\documentclass[twocolumn]{IEEEtran}
\usepackage[usenames,dvipsnames]{color}
\usepackage{subfigure}
\usepackage{graphicx}
\usepackage{amsmath}
\usepackage{color}
\usepackage{multicol}
\usepackage{amssymb}
\usepackage{balance}

\usepackage{amsfonts}%
\usepackage{verbatim}%
\usepackage{enumerate}%
\usepackage{psfrag}
\usepackage{epsfig}
\usepackage{multicol}
\usepackage{setspace}
\usepackage{dsfont}

\usepackage{algorithm}
\usepackage{algorithmic}
\usepackage{cite}
\usepackage{float}
    
\begin{document}

\title{Optimal Cooperative Cognitive Relaying and Spectrum Access for an Energy Harvesting Cognitive Radio: Reinforcement Learning Approach}
\author{\IEEEauthorblockN{Ahmed El Shafie$^{*\ddag}$,  Tamer Khattab$^\ddag$, Hussien Saad$^{*}$, Amr Mohamed$^\dag$}

\IEEEauthorblockA{$^\dag$ Computer Science and Engineering Dept., Qatar University, Doha, Qatar.\\
$^*$ Wireless Intelligent Networks Center (WINC), Nile University, Giza, Egypt.\\
$^\ddag$ Electrical Engineering Dept., Qatar University, Doha, Qatar.\\
}
}
\date{}
\maketitle
\thispagestyle{empty}

\begin{abstract}
In this paper, we consider a cognitive setting under the context of cooperative communications, where the cognitive radio (CR) user is assumed to be a self-organized relay for the network. The CR user and the PU are assumed to be energy harvesters. The CR user cooperatively relays some of the undelivered packets of the primary user (PU). Specifically, the CR user stores a fraction of the undelivered primary packets in a relaying queue (buffer). It manages the flow of the undelivered primary packets to its relaying queue using the appropriate actions over time slots. Moreover, it has the decision of choosing the used queue for channel accessing at idle time slots (slots where the PU's queue is empty). It is assumed that one data packet transmission dissipates one energy packet. The optimal policy changes according to the primary and CR users arrival rates to the data and energy queues as well as the channels connectivity. The CR user saves energy for the PU by taking the responsibility of relaying the undelivered primary packets. It optimally organizes its own energy packets to maximize its payoff as time progresses.
\end{abstract}
\section{Introduction}
  \IEEEPARstart{S}\small{econdary} usage of the licensed frequency bands can efficiently improve the spectral density of the under-utilized licensed spectrum. Cognitive radio (CR) users are intelligent terminals that use cognitive technologies to be fully aware of the environmental variations. A CR user should exploit methodologies of learning and reasoning to dynamically reconfigure its communication parameters.

Cooperative diversity, which  is a recently emerging technique
for wireless communications, has obtained a wide attention recently. Cooperative cognitive relaying, which involves cooperation between primary and secondary nodes in cognitive radio networks, has been investigated in many existing works, e.g., \cite{stability_cog1,khattab,erph}. In \cite{stability_cog1}, the authors investigate a cognitive network with one primary user (PU) and one CR user. The cognitive terminal optimally adjusts its power such that the secondary queue mean service rate is maximized while maintaining all queues in the network stable. In \cite{khattab}, the authors
 consider that the CR terminal can use the primary spectrum when the PU is inactive under a priority in transmission assigned to the relaying queue.
 The CR user admits a predefined fraction of the undelivered packets of the PU to be relayed. The authors optimize over that fraction to achieve the minimum secondary queueing delay.

Energy harvesting technology has been recently incorporated to the transmitting terminals of wireless networks. Optimal energy management has been addressed in many
papers such as \cite{sharma2010optimal,ho2010optimal,ozel2011transmission}. The authors of \cite{sharma2010optimal}, Sharma \emph{et al.}, obtained the optimal
energy management policies for an energy harvester.
In \cite{ho2010optimal}, energy allocation over a finite horizon is
considered with the objective of maximizing the throughput
and taking into account time-varying channel conditions.
In
\cite{ozel2011transmission}, communication by an energy harvester over
a wireless fading channel is considered. Stochastic dynamic
programming is used to solve for the optimal online policy that maximizes the average number of bits delivered by a deadline
under stochastic fading and energy arrival processes with
causal channel state feedback.

In a cognitive setting, there are several works that include energy harvesting transmitters, e.g., \cite{park2011energy,hoang2009opportunistic,pappas2012optimal,Sult1210:Optimal,krikidis2012stability,ourletter,ElSh1312:Optimal}. In \cite{park2011energy},
a Markov decision process (MDP) is proposed to obtain the
optimal secondary access policy under perfect spectrum sensing.
The authors of \cite{hoang2009opportunistic} investigate an energy constrained
cognitive terminal without explicitly involving an
energy queue.
The authors of \cite{pappas2012optimal} investigate a
scenario with one rechargeable PU and one cognitive terminal. The maximum stable-throughput region is characterized. In \cite{Sult1210:Optimal}, the authors investigate the maximum stable secondary mean service rate under the stability of the primary and secondary queues and with MPR capability added to the receiving nodes. The network model consists of a PU and an energy harvesting CR user. In \cite{krikidis2012stability}, Krikidis \emph{et al.} investigate the impact
of cooperation in a three-node network with energy harvesting
nodes and bursty data traffic from network layer standpoint. The authors derive the stability region of the system as well as
the required transmitted power for both a non-cooperative and an
orthogonal decode-and-forward cooperative protocols. In \cite{ourletter}, the authors assume a simple access scheme where the SU randomly accesses the channel at the beginning of the time slot without performing channel sensing to exploit the MPR capability of the receiving nodes. The maximum throughput of a saturated SU is obtained under stability and queueing delay constraints on the primary queue.
 In \cite{ElSh1312:Optimal}, the authors propose a cognitive setting with one energy harvesting PU and one energy harvesting SU. The SU randomly selects a sensing duration from a predefined set to discern the primary activity. The authors obtain the maximum stable-throughput of the SU under stability of the PU's queue. In \cite{wcmpaper}, El Shafie \emph{et al.} investigate the impact of cooperation from a network layer point of view on a network composing of an energy harvesting SU and a PU plugged to a reliable power supply. The SU utilizes the spectrum whenever the PU is inactive. The authors assume that one energy packet is dissipated in either data decoding or data transmission. Due to the interaction of queues, inner and outer bounds are derived for the secondary throughput.

In this paper, we assume that the CR user senses the channel for $\tau$ seconds from the beginning of the time slot to detect the activity of the PU. Based on the sensed primary state, the SU has to take an action. Thus, the action is taken after $\tau$ seconds from the beginning of the time slot; exactly after sensing the channel. If the PU is sensed to be inactive, the CR user has to choose between being idle to the end of the time slot or transmitting a packet either from its own packets or from the relaying packets. If the PU is active, the CR user has to choose between being idle or accepting the primary packet. Unlike most of the existing works, we do not assume a decoupled M/D/1\footnote{The notation of discrete-time M/D/1 queue is used to describe a queueing
system with Bernoulli arrival process and deterministic service process.} with unity service rate model for the energy queues (for further details, the reader is referred to \cite{pappas2012optimal} and \cite{Sult1210:Optimal} and the references therein), which is a trivial model and provides an inner bound on the performance. Moreover, it makes the queue capacity useless as shown in \cite{commentfile}. Moreover, we assume a finite length energy and data queues. We do not consider either dominance system approach or always nonempty queues to decouple the queues as proposed in many works, e.g., \cite{pappas2012optimal,Sult1210:Optimal,sadek}. Furthermore, in contrast to the conventional modeling of the arrival processes, where the arrival processes are assumed to be independent and identically distributed Bernoulli random processes~\cite{pappas2012optimal,Sult1210:Optimal,wcmpaper}, we assume correlated arrivals at each queue and model the arrival processes of the queues as {\it Markov modulated Bernoulli processes}. The proposed approach and the analysis presented in this paper are generic and can be applied to any system.

\section{SYSTEM MODEL}\label{sysmod}
   \begin{figure}
\normalcolor
\centering
  \includegraphics[width=1\columnwidth]{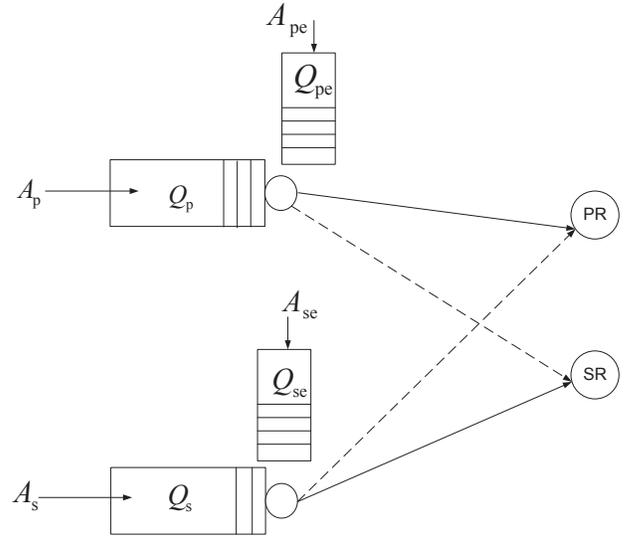}
  \caption{Primary and secondary links and queues. The solid lines are the communication channels, while the dotted lines are the interference channels. The primary and secondary receivers are denoted by PR and SR, respectively. Note that the number of arrivals at $Q_k$ in time slot $t$ is $A_k$.}\label{relayednw}
\end{figure}
The network model adopted in this paper composes of two energy harvesters sharing the same channel resources with different priority of channel accessing. The PU is the user with the highest priority and accesses the channel unconditionally, whereas the CR user is the lowest priority user and accesses the channel whenever the PU is declared to be inactive. The inactivity of the PU occurs due to the lack of either the energy packets in its energy queue or/and data packets in its data queue. The network model is depicted in Fig. \ref{relayednw}.

\begin{figure}
\normalcolor
  \includegraphics[width=1\columnwidth]{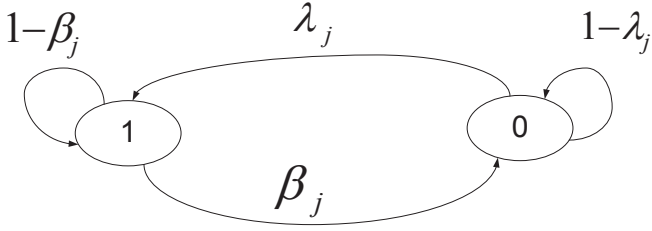}
  \caption{Two state Markov model of Markov modulated Bernoulli process for queue $Q_k$, $k\in\{\rm p,pe,s,se\}$.}\label{forqueues}
\end{figure}

We assume that the PU has two different types of buffers; a data buffer to store its incoming data packets, denoted as $Q_{\rm p} $,
and an energy buffer to store the energy packets (tokens) harvested from the environment, denoted as $Q_{\rm pe} $. The CR user has three buffers; $Q_{\rm s} $ to store its
own arrived data traffic, $Q_{\rm ps} $ to store the accepted primary packets for relaying, and $Q_{\rm se} $ to store the harvested energy packets from the environment. We assume that all buffers are
{\bf finite} length. Precisely, queue $Q_j$, $j\in\{\rm p,pe,s,ps,se\}$, can maintain at most $\mathcal{B}_j$ packets. We consider a time-slotted transmissions. The duration of one slot is $T$ seconds. All packets have
the same size and each contains $b$ bits. The packets arrival processes to the PU and SU queues are assumed to be Markov modulated Bernoulli processes \cite{ozekici1997markov} where the probability of arrival occurrence of a Bernoulli process evolves over time according to a Markov chain. The arrivals at each queue are assumed to respect the following two state Markov chain (shown in Fig. \ref{forqueues}):
$$\Bigg(\begin{array}{cc}
  1-\lambda_{k} & \lambda_{k}  \\
  \beta_{k} & 1-\beta_{k}
\end{array}\Bigg)$$
where $\lambda_{k}$ denotes the probability of having no arrived packet at queue $Q_k$, $k\in\{\rm p,pe,s,se\}$, in time slot $t+1$ when there was no arrived packet in time slot $t$ and $\beta_{k}$ denotes the probability of having no arrived packet at queue $Q_k$ in time slot $t+1$ when there was an arrived packet in time slot $t$. We assume that arrivals are independent random variables from queue to queue. We denote the number of arrivals at $Q_k$ by $A^t_k$ where $A^t_k=1$ if there is an arrived packet in time slot $t$ and zero otherwise.

%

\begin{figure}[t]
\normalcolor
  \includegraphics[width=1\columnwidth]{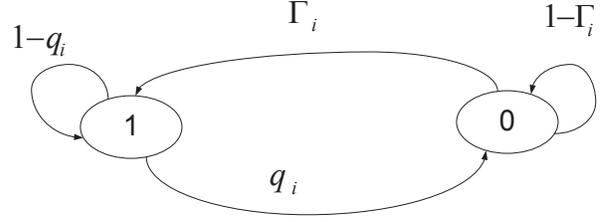}
  \caption{Two state Markov model for link $i$.}\label{forchannels}
\end{figure}

%
The radio channel gain of the links between any pair of
nodes $h^t_i$ is assumed to be zero mean circularly symmetric
complex Gaussian random variable with variance $\sigma_{i}^2$, i.e., $\mathcal{CN}(0,\sigma_i^2)$, and independent for all $i$, where $i$ reads `${\rm p}$' for the
primary link, `${\rm s}$' for the secondary link, `${\rm ps}$' for the link between
the PU and the CR user, and `${\rm sp}$' for the
link between the CR user and the primary destination.
Each link is perturbed by a thermal noise which is modeled as complex additive white Gaussian noise (AWGN) with zero mean and variance $\mathcal{N}_\circ$ and independent for
all links. We assume a two state Markov channels. Specifically, the $i$th link follows (shown in Fig. \ref{forchannels}):
$$\Bigg(\begin{array}{cc}
  1-\Gamma_i & \Gamma_i  \\
  q_i & 1-q_i
\end{array}\Bigg)$$
where $\Gamma_i$ is the probability of link $i$ being not in outage in time $t+1$ given that it was in outage in time slot $t$ and $ q_i$ is the probability of the link being in outage in time $t+1$ given that it was not in outage in time slot $t$.

 Let $\overline{\mathcal{X}}=1-\mathcal{X}$, $1\big[\mathcal{F}\big]=1$ if the event $\mathcal{F}$ is true, and ${\rm I}^{t}_{c_{i}}$ be the indication of the channel state and is equal to unity if link $i$ is connected and zero otherwise. We consider that the channel is ON (connected) if the transmitted rate is less than or equal to the channel capacity; otherwise, it is OFF (disconnected). We assume that the SU knows the channels gains perfectly at the beginning of the time slot. The primary channel can be sent from the primary destination over a dedicated narrow-band during the sensing phase of the spectrum.\footnote{We would emphasize here that the proposed protocol is based on the cooperation between users. Thus, the primary system aids the secondary system for increasing the performance of the system.}

The medium access control is assumed to obey the following rules.
\begin{itemize}
\item At the beginning of the time slot, the CR user senses the channel for $\tau$ seconds from the beginning of the time slot to declare the state of activity of the PU.\footnote{The sensing duration should be large enough for channels status estimation and perfect channel sensing. Note that the channels activities can be sent using one-bit feedback signal from nodes to SU. The nodes need only to send the state of the channel, i.e., ON or OFF. This can be sent during the sensing duration either sequentially or at the same time using different narrow-band frequency bands for each node.}
    \item The sensing process outcome is recorded as a binary value at the secondary terminal. In particular, it is recorded as `$1$' if the PU is active or `$0$' if the PU is inactive.
    \item If the PU is sensed to be inactive, the CR user has to choose between being idle till the end of the time slot or transmitting a packet either from its own queue, $Q_{\rm s} $, or from the relaying queue, $Q_{\rm ps}$.
        \item If the PU is active, the CR user has to choose between being idle till the end of the time slot or accepting the primary packet.
\item If the primary destination could not decode the PU packet correctly and the CR user could decode and decide to accept the packet, the SU has to send acknowledgement/negative-acknowledgment (ACK/NACK) message to the PU based on the result decoding of the packet.
These packets are then dropped from the primary queue.
\item In case both the CR user and the primary destination fail
to decode the primary data, a retransmission of the packet is
initiated by the PU at the following time slots.\footnote{If the PU receives at least one ACK in a time slot, it drops the packet from its queue. If the PU receives two NACKs, it retransmits the packet at the following time slots.}
\end{itemize}

 We assume that the overhead for transmitting the ACKs and NACKs is negligible relative to packet sizes. The second assumption we make is that the errors in packet acknowledgement feedback are negligible. This assumption
is reasonable for short length ACK/NACK packets as low rate and strong
codes can be employed in the feedback channel \cite{sadek}. In addition, nodes cannot transmit and receive at the same time which is a common assumption where terminals are equipped with single transceivers \cite{erph}.

According to the previous description, the SU has four distinct actions. After $\tau$ seconds from the beginning of the time slot, the SU has to select one of the possible actions. Note that the CR user should optimally distribute its energy packets among the transmissions of the data packets to achieve the highest possible performance.

\section{Queues Arrival and Service Processes}\label{serviceprocess}
As mentioned earlier, the CR user has four possible actions. The set of actions is $\mathcal{A}\!=\! \{a_1,a_2,a_3,a_4\}$, where $a_1$: transmitting a packet from $Q_{\rm s} $, $a_2$: transmitting a packet from $Q_{\rm ps} $, $a_3$: accepting a packet from the PU, and $a_4$: remaining idle (CR user is idle). Note that the optimal action vector in a given time slot satisfies the following constraint:
 \begin{equation}
 \sum_{n=1}^4 a^t_n=1, \ \forall \ t=0,T,2T,\dots
 \end{equation}
 This condition means that there is only one action per time slot.

  A packet from $Q_{\rm s} $ is served if the CR user has energy packets in its energy queue, the CR user accesses the channel using $Q_{\rm s} $, the channel to the respective receiver is ON, and the PU is inactive. Mathematically, the service process of $Q_{\rm s}$ can be modeled as:
\begin{equation}
\mathcal{R}^t_{\rm s} =a_1{\rm I}^{t}_{c_{\rm s} }\bigg(1-{\rm I}^t_{Q_{\rm p} }{\rm I}^t_{Q_{\rm pe} }\bigg){\rm I}^t_{Q_{\rm se} }
\end{equation}
 The term ${\rm I}^t_{Q_{j}}$ equals to unity if the queue $Q_j$ is not empty and zero if the queue is empty. Note that the PU is active if both its data and energy queues are nonempty, i.e., ${\rm I}^t_{Q_{\rm p} }{\rm I}^t_{Q_{\rm pe} }=1$. Thus, the term $1-{\rm I}^t_{Q_{\rm p} }{\rm I}^t_{Q_{\rm pe} }$ indicates the inactivity of the PU.

Consider the relaying queue. A packet departs $Q_{\rm ps} $ if the CR user has energy in its energy queue, the CR user decides to access the channel with a packet from $Q_{\rm ps} $, the channel to the respective receiver is ON, and the PU is inactive. Mathematically, the service process can be modeled as:
\begin{equation}
\mathcal{R}^t_{\rm ps} =a_2{\rm I}^{t}_{c_{\rm sp} }\bigg(1-{\rm I}^t_{Q_{\rm p} }{\rm I}^t_{Q_{\rm pe} }\bigg){\rm I}^t_{Q_{\rm se} }
\end{equation}

The arrival process to the relaying queue is described as follows. A packet is arrived to the relaying queue if the primary queue is nonempty, the relaying queue is not full, the channel between the PU and its respective receiver is OFF, the channel from the PU to the SU is ON, and the SU decides to accept the packet. Mathematically, the arrival process to $Q_{\rm ps}$ is given by
\begin{equation}
A^t_{\rm ps} =a_3 {\rm I}^{t}_{c_{\rm ps} }{\rm I}^{t}_{Q_{\rm p} }{\rm I}^t_{Q_{\rm pe} }\overline{{\rm I}^{t}_{c_{\rm p} }} 1[Q_{\rm ps}<\mathcal{B}_{\rm ps}]
\end{equation}

A packet from the secondary energy queue is consumed in either one of the following events. If the SU accesses the channel either from its data queue or from the relaying queue. Mathematically, the process is given by
\begin{equation}
\mathcal{R}^t_{\rm se} =a_1 {\rm I}^t_{Q_{\rm s} }+a_{2}{\rm I}^t_{Q_{\rm ps} }
\end{equation}

  Given that the PU has energy in its energy queue, a packet from the PU's data queue is served in either one of the following events. If the channel between the PU and its respective receiver is ON and the SU remains idle; or if the channel between the PU and its respective destination is OFF, the channel between the PU and the SU is ON, the relaying queue is not full, and the SU decides to accept the packet. The process is modeled as follows:
\begin{equation}
\begin{split}
    \mathcal{R}^t_{\rm p} &={\rm I}^t_{Q_{\rm pe} }\bigg(a_4{\rm I}^{t}_{c_{\rm p} }\!+\!a_3{\rm I}^{t}_{c_{\rm ps} }\overline{{\rm I}^{t}_{c_{\rm p} }} 1[Q_{\rm ps}<\mathcal{B}_{\rm ps}]\bigg)
    \end{split}
\end{equation}

Since the PU accesses the channel unconditionally whenever it has energy and data packets, a packet from the PU energy queue is consumed if the primary queue is nonempty. That is,
\begin{equation}
\mathcal{R}^t_{\rm pe} ={\rm I}^t_{Q_{\rm p} }
\end{equation}

We assume that
departures occur before arrivals, and the queue size is measured
at the beginning of the slot \cite{rao1988stability}. The evolution of queue $Q_j$ is then given by
\begin{equation}\label{5}
\begin{split}
    Q^{t+1}_j&\!=\!\min\Bigg\{\!\max\biggr\{\!Q^{t}_{j}\!-\!\mathcal{R}^t_{j},0\!\biggr\}\!+\!A^{t}_{j},\mathcal{B}_j\!\Bigg\}, j\in\{\!\rm p,pe,s,ps,se\!\}
    \end{split}
\end{equation}
where $\max\{\cdot,\cdot\}$ and $\min\{\cdot,\cdot\}$ return the maximum and the minimum among the values in the argument, respectively.
\section{{\it Q}-learning Algorithm}\label{Qlear}
 The prime goal in the reinforcement learning (RL) is to choose actions over time so as to maximize the expected value of the total payoff of the learner (agent or user). The CR user will be able to achieve the adaptive optimal policy according to the mean arrival rates of the queues and outage probabilities of all channels in order to maximize its expected payoff as time progresses. MDPs are considered powerful frameworks for solving problems of sequential decision making under uncertainty \cite{mitchell1997machine,sutton1998reinforcement,sutton1999between}.
Bellman's equation, which forms the foundation for many dynamic programming approaches
to solving MDPs, is given by:
\begin{equation}\label{bellman}
    V(s)=\mathcal{R}(s,a)+\gamma \sum_{\hat {s} \in \mathcal{S}} P({\hat {s}}|s,a) V(\hat {s})
\end{equation}
where $V(s)$ is the discounted cumulative reward and $\gamma$ is a constant that determines the relative value of delayed versus immediate rewards. Choosing the discount factor $\gamma$ smaller than $1$ ensures convergence of the sum. For every state $s$ we may
investigate what the best policy (action) is, and what its value would be. Let us define the optimal value function as
the maximum value function among all value functions, it satisfies the Bellman equation, and is given by
\begin{equation}\label{bellman2}
    V^*(s)=\max_{a} \bigg[\mathcal{R}(s,a)+\gamma \sum_{{\hat {s}} \in \mathcal{S}} P(\hat {s}|s,a) V(\hat {s})\bigg]
\end{equation}
where $V^*(s)$ gives the maximum discounted cumulative reward that the
agent can obtain starting from state $s$; that is, the discounted cumulative reward
obtained by following the optimal policy beginning at state $s$ \cite{mitchell1997machine}. The policy is a function that maps the state space to action space, i.e., $\pi: \mathcal{S} \rightarrow A$. The optimal policy is given by:
\begin{equation}\label{bellman3}
   \pi^*(s)=\underset{a}{\rm argmax} \bigg[\mathcal{R}(s,a)+\gamma \sum_{{\hat {s}} \in \mathcal{S}} P(\hat {s}|s,a) V(\hat {s})\bigg]
\end{equation}
The reward function is defined according to the states and actions and it aims at maximizing the weighted sum of the service rates of the CR user queues subject to some predefined constraints. Mathematically, the immediate reward function is given by
\begin{equation}
\begin{split}
\mathcal{R}(s,a)=\omega \mathcal{R}_s {\rm I}_{Q_{\rm s}}&+(1-\omega) \mathcal{R}_{\rm ps}{\rm I}_{Q_{\rm ps}} \\& -\mathcal{K} \Bigg[{\rm I}_{Q_{\rm p}}{\rm I}_{Q_{\rm pe}}(a_1+a_2)+a_1\overline{{\rm I}_{c_{\rm s}}{\rm I}_{Q_{\rm s}}{\rm I}_{Q_{\rm se}}}\\&  \,\,\,\,\,\,\,\,\,\,\,\,\,\,\ +a_2\overline{{\rm I}_{c_{\rm ps}} {\rm I}_{Q_{\rm ps}}{\rm I}_{Q_{\rm se}}}\!+\!a_3 1[Q_{\rm ps}\!=\!\mathcal{B}_{\rm ps}]\!\\&  \,\,\,\,\,\,\,\,\,\,\,\,\,\,\ +\!a_3({\rm I}_{c_{\rm p}}{\rm I}_{Q_{\rm p}}{\rm I}_{Q_{\rm pe}}+\overline{{\rm I}_{c_{\rm ps}} {\rm I}_{Q_{\rm p}}{\rm I}_{Q_{\rm pe}}})\Bigg]
\end{split}
\end{equation}
where $\omega$ is a fixed number that belongs to the set $[0,1]$ and $\mathcal{K}$ is a penalty constant. The rational behind this cost function
is that the CR users cannot transmit at the same time with the PU to avoid a sure collision event, which is specified by $-\mathcal{K} {\rm I}_{Q_{\rm p}}{\rm I}_{Q_{\rm pe}}(a_1+a_2)$; to avoid wasting the secondary energy when channels are in outage, which is specified by  $-\mathcal{K}(a_1(\overline{{\rm I}_{c_{\rm s}}{\rm I}_{Q_{\rm s}}{\rm I}_{Q_{\rm se}}})+a_2\overline{{\rm I}_{c_{\rm ps}} {\rm I}_{Q_{\rm ps}}{\rm I}_{Q_{\rm se}}})$; to avoid decoding the primary packet when the relaying queue is full, which is specified by $a_3 1[Q_{\rm ps}=\mathcal{B}_{\rm ps}]$; and to avoid using an action when the corresponding queue is empty or the secondary energy queue is empty or to take packet acceptance action when the PU is inactive. Note that, the more $\omega$ indicates more emphasizing on the service rate of $Q_{\rm s}$ (secondary throughput), and the lower the $\omega$ the more emphasizing on the service rate of $Q_{\rm ps}$.

In {\it Q-learning}, the agent, which is the CR user in this work, interacts with the environment to obtain the consecutive actions that maximize the accumulative payoff of the weighted sum of the secondary queues, $Q_{\rm s} $ and $Q_{\rm ps} $, mean service rates. In particular, the CR user aims at maximizing the expected weighted sum of the its queue service rates. It is assumed that the environment is a finite-state discrete time
stochastic dynamical system.

\begin{algorithm}[t]{}
\caption{{\it Q-learning} algorithm}
\begin{algorithmic}\label{algo1}
\STATE{Initialize:}
\STATE{let $t=0$}
\FOR{ each $s \in \mathcal{S}$ and $a \in \mathcal{A}$}
\STATE{initialize the {\it Q} value}
\ENDFOR
\STATE{Initialize $s^t$}
\STATE{Learning:}
\LOOP
\STATE{generate a random number $\ell$ between $0$ and $1$}
\IF {$\ell<\mu$}
\STATE{select one of the actions randomly}
\ELSE
\STATE{select the action $a^t$ characterized by the \\ \,\,\ maximum {\it Q-value}}
\ENDIF
\STATE{execute $a^t$}
\STATE{receive an immediate reward $\mathcal{R}(s^t,a^t)$}
\STATE{observe the next state $s^{t+1}$}
\STATE{update the table entry as follows:}
\STATE{$s^t\leftarrow s^{t+1}$}
    \STATE{$\mathcal{Q}(s,a)\!\leftarrow \! \mathcal{Q}(s,a)\!+\!\alpha\bigg(\mathcal{R}(s,a)\!+\!\gamma \underset{\hat{a}}{\max}\mathcal{Q}(\hat{s},\hat{a})\!-\!\mathcal{Q}(s,a)\bigg)$}
\ENDLOOP
\end{algorithmic}
\end{algorithm}

The interactions between the
CR user and the environment at every time slot
$t$ is described as follows.
\begin{itemize}
                             \item The CR user senses the channel for $\tau$ seconds.
                              \item The CR user observes its state $s$.
                              \item Based on $s$, the CR user chooses an action $a$ from the feasible actions set $\mathcal{A}$.
                              \item The CR user receives an immediate reward $\mathcal{R}(s,a)$.
                              \item A transition to the state $\hat{s}$ takes place.
                              \item The learning process is repeated until convergence to the optimal policy.
                            \end{itemize}
 The {\it Q-learning} algorithm (Algorithm \ref{algo1}) is the most popular powerful and widely used form of reinforcement learning due to the naive implementation of this method. It obtains the optimal {\it Q-values}, rather than state-values. The update rule for {\it Q-learning} is
\begin{equation}\label{Qvalue}
    \mathcal{Q}(s,a)\leftarrow \mathcal{Q}(s,a)+\alpha\biggr[\mathcal{R}(s,a)+\gamma \max_{\hat{a}}\mathcal{Q}(\hat{s},\hat{a})-\mathcal{Q}(s,a)\biggr]
\end{equation}
where $\alpha$ is the learning rate and $\gamma$ is the discount factor. The idea of update rule is that the part $\mathcal{R}(s,a)+ \gamma \underset{\hat{a}}{\max} \mathcal{Q}(\hat{s},\hat{a})$ is an estimate of the {\it Q-value} $\mathcal{Q}(s,a)$. Watkins proved that this method will converge to the {\it Q-values} for the optimal policy, $\mathcal{Q}^*(s,a)$, if two conditions were met, every state-action pair has to be visited infinitely often and the learning rate $\alpha$ decays over time. A proof of convergence for {\it Q-learning} based on that outlined in Watkins was presented in \cite{wat2}. The authors show that {\it Q-learning} converges to the optimum action-values with probability $1$ so long as all actions
are repeatedly sampled in all states and the action-values are represented discretely. The objective of the CR user is to find an optimal policy
$\pi^*(s) \in \mathcal{A}$  for each state $s$, to maximize some cumulative measure
of the cost $\mathcal{R}(s,a)$ received over time. We define the evaluation function,
denoted by $\mathcal{Q}(s, a)$, as the expected total discount cost over an infinite time and it is given by
\begin{equation}\label{Qvalue2}
    \mathcal{Q}(s,a)=\mathcal{E}\biggr\{\sum_{t=0}^{\infty}\gamma \mathcal{R}(s,\pi(s))|s_0=s\biggr\}
\end{equation}
where $\mathcal{E}\big\{\cdot\big\}$ denotes the expected value. If the selected action $a$ in time slot $t$ following the policy $\pi(s)$ which is corresponding to the optimal
policy $\pi^*(s)$, the {\it Q-function} is maximized with respect to the
current state. It can be shown that (\ref{Qvalue2}) is given by
\begin{eqnarray}
    \mathcal{Q}(s,a)=\mathcal{E}\biggr\{\mathcal{R}(s,a)\biggr\}+\gamma \sum_{\hat{s} \in \mathcal{S}}P(\hat{s}|s,a)\mathcal{Q}(\hat{s},a)
    \label{eqnQ}
\end{eqnarray}
Recall that $P(\hat{s}|s,a)$ is the transition probability from state $s$ to next
state $\hat{s}$, when action $a$ is executed. Eqn. (\ref{eqnQ}) indicates that the Q-function of the current
state-action pair, can be represented in terms of the expected
immediate cost of the current state-action pair and the {\it Q-value}
of the next state-action pairs. {\it Q-learning} aims at determining an optimal stationary policy $\pi(s)$, without knowing $\mathcal{E}\{\mathcal{R}(s,a)\}$ and $P(\hat{s}|s,a)$. The states are defined as follows. Without loss of generality, we divide the CR user's queues to $\mathcal{N}$ portions. In particular, each queue in the CR terminal is divided to $\mathcal{N}$ portions as follows:
\begin{equation}
\mathcal{L}(Q_n)= \left\{ \begin{array}{ll}
         0 & \mbox{if $  Q_n = 0 $}\\
         1 & \mbox{if $ 0< Q_n \le \nu_{n,th,1} $}\\
        2 & \mbox{if  $\nu_{n,th,1} +1\le Q_n \le \nu_{n,th,2}$ }\\
        3 & \mbox{if  $\nu_{n,th,2} +1\le Q_n \le \nu_{n,th,3}$ }\\ \vdots  & \mbox{  $\vdots$ } \\
        \mathcal{N}\!-\!1 & \mbox{if  $Q_n \ge \nu_{n,th,\mathcal{N}-2}+1$ }
        \end{array} \right.
        \label{partitioning}
        \end{equation}
        where $n\in \{\rm s,ps,se\}$ and $\nu_{n,th,h}$ is the $h$th threshold of the queue $Q_n$.

%
        The state vector, at any time instant $t$, is formed as
        \begin{equation}
        \mathcal{S}^t\!=\!\bigg[{\rm I}^t_{Q_{\rm p} }{\rm I}^t_{Q_{\rm pe} },\mathcal{L}(Q^t_{\rm ps} ),\mathcal{L}(Q^t_{\rm se} ),\mathcal{L}(Q^t_{\rm s} ),{\rm I}^t_{\rm c,sp},{\rm I}^t_{\rm c,s},{\rm I}^t_{\rm c,p},{\rm I}^t_{\rm c,ps}\bigg]
        \end{equation}
where ${\rm I}^t_{Q_{\rm p} }{\rm I}^t_{Q_{\rm pe} }$ represents the activity of the PU and is ascertained from channel sensing. According to the above description, the total number of states is $2^5 \times \mathcal{N}^3$, where $2$ represents the possibility of the binary valued channels states.

With respect to the {\it Q-learning} algorithm, the learning rate is
$\alpha= 0.5$ and the discount factor is $\gamma = 0.9$. We also introduce
a probability $\mu = 0.05$ of visiting random states in the initial
$60\%$ of the {\it Q-learning} iterations. This parameter is used
in the action selection procedure to guarantee that the final
policy is a global optimum and not a local one \cite{Ana}.
\section{Results, Simulations and Conclusions}\label{num}
In this section, we provide some simulations of the system. Simulations are executed using $\beta_{\rm p}\!=\!\lambda_{\rm p}~\in~[0,1]$. Let `CS' denote the cooperative system and `NC' denote the non-cooperative system. We assume that each buffer, in the network, is of size $20$ packets. We split each queue of the CR user to $\mathcal{N}=4$ portions which leads to the availability of $\mathcal{N}^3=64$ states. The thresholds of the queues are: $\nu_{n,th,0}=6$ and $\nu_{n,th,1}=12$ for all $n$. The capacity of all queues belonging to the system is $\mathcal{B}_j=\mathcal{B}=20$ packets, for all $j$, packets. The rest of the parameters are: $\mathcal{K}=10$, $\Gamma_{\rm p}=0.2$,
$q_{\rm p}=0.4$,
$\Gamma_{\rm s}=0.6$,
$q_{\rm s}=0.1$,
$\Gamma_{\rm ps}=0.7$,
$q_{\rm ps}=0.2$,
$\Gamma_{\rm sp}=0.8$,
$q_{\rm sp}=0.05$,
$\lambda_{\rm s}=0.4$,
$\beta_{\rm s}=0.4$,
$\lambda_{\rm se}=0.8$,
$\beta_{\rm se}=0.4$,
$\lambda_{\rm pe}=0.4$, $\beta_{\rm pe}=0.4$ and $\beta_{\rm p}=\lambda_{\rm p}$.

\begin{figure}
  \centering
  \includegraphics[width=1\columnwidth]{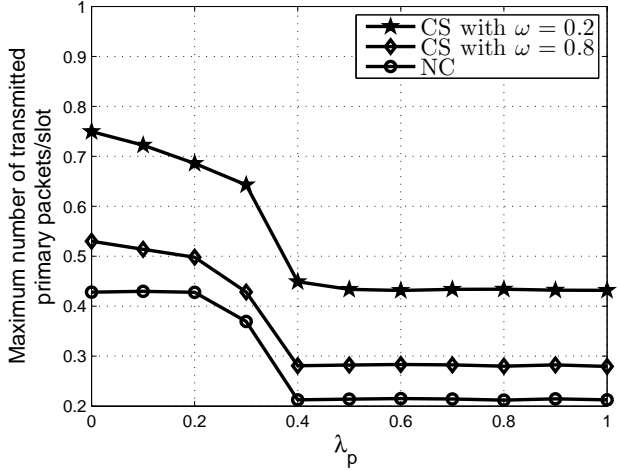}\\
  \caption{The maximum primary transmitted packets per time slot.
  }\label{fig1}
  \end{figure}

\begin{figure}
  \centering
  \includegraphics[width=1\columnwidth]{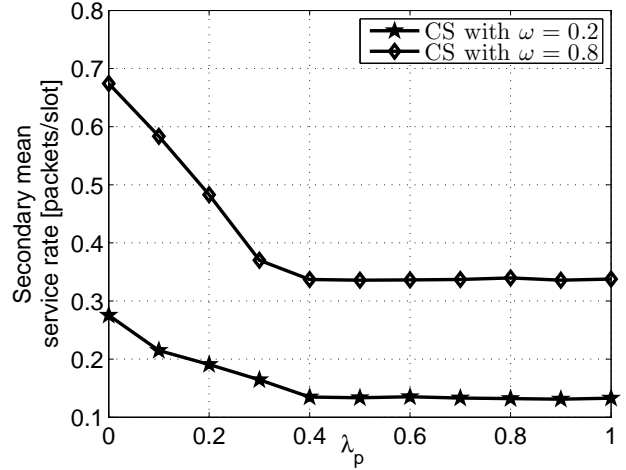}\\
  \caption{The maximum secondary service rate in packets/slot.
  }\label{fig2}
  \end{figure}

As shown in Fig. \ref{fig1}, the primary maximum number of transmitted packets per time slot increases with cooperation. Moreover, decreasing $\omega$ increases the service rate of the relaying queue; hence, increases the primary transmitted packets per time slot. The beneficial gain of cooperation is shown in the figure. Fig. \ref{fig2} demonstrates the maximum mean service rate of the secondary own data queue. As seen from the figure, increasing $\omega$ emphasizes on the secondary service; hence, increases the secondary mean service rate. From the figures, we conclude that the cooperation is important for both users and it boosts their throughputs. Furthermore, The parameter $\omega$ manages the throughputs of users and it can be used to archive certain quality of service for each user.

In the paper, we have investigated a cooperative energy harvesting CR user sharing the channel resources with a PU. The CR user has to decide on taking a specific action preceded by a channel sensing from the beginning of every time slot. The action taken at each state is selected to, on the average, maximize the secondary expected utility as time progresses. Unlike most of the existing work, we have considered finite queue lengths and characterized the system performance with the existence of strong queue interaction. We also have considered Markov modulated Bernoulli arrival processes at queues. The optimal policy has been obtained using {\it Q-learning} algorithm where each state is assigned an action.
\bibliographystyle{IEEEtran}
\bibliography{IEEEabrv,nw}
\balance
\end{document}